\documentclass{article}
\usepackage{spconf,amsmath,graphicx}
 \usepackage{booktabs}
 \usepackage[thinc]{esdiff}
 \usepackage{multirow,xcolor}
 \usepackage{subfig}
\newcommand{\be}{\begin{eqnarray}}
\newcommand{\ee}{\end{eqnarray}}

\usepackage{comment}
\usepackage{multirow}
\usepackage{graphicx}
\usepackage{url}
\usepackage{xcolor,colortbl}
\usepackage{enumitem}
\usepackage{MnSymbol,bbding,pifont}
\newcommand{\method}{ZEST}
\newcommand{\methodexp}{zero shot emotion transfer }

\definecolor{Gray}{gray}{0.85}
\definecolor{LightCyan}{rgb}{0.88,1,1}

\title{Zero Shot Audio to Audio Emotion Transfer With Speaker Disentanglement }
\name{Soumya Dutta and Sriram Ganapathy{\thanks{This work was partly funded by   grants provided by British Telecom.}}}

\address{
  Learning and Extraction of Acoustic Patterns (LEAP) lab, Indian Institute of Science, Bangalore. \\
  E-mail - {\{soumyadutta,sriramg\}@iisc.ac.in}
  }
\ninept 

\begin{document}

\maketitle
\begin{abstract}
The problem of audio-to-audio (A2A) style transfer involves replacing  the style features of the source audio with those from the target audio while preserving the content related attributes of the source audio.
In this paper, we propose an efficient approach, termed as {\textbf{Z}ero-shot  \textbf{E}motion \textbf{S}tyle \textbf{T}ransfer (\textbf{\method})},  that allows the transfer of emotional content present in the given source audio with the one embedded in the target audio while retaining the speaker and speech content from the source. The proposed system builds upon decomposing speech into semantic tokens, speaker representations and emotion embeddings. 
 Using these factors, we propose a framework to reconstruct the pitch contour of the given speech signal and train a decoder that reconstructs the speech signal. 
The model is trained using a self-supervision based reconstruction loss. During conversion, the emotion embedding is alone derived from the target audio, while rest of the factors are derived from the source audio. In our experiments, we show that, even without using parallel training data or labels from the source or target audio, we illustrate zero shot emotion transfer capabilities of the proposed \method ~model using objective and subjective quality evaluations.
\end{abstract}


\section{Introduction}
Artificial emotional intelligence \cite{schuller2018age} encompasses methods that enable machines to understand and interact with human expressions of emotions. 
The style transfer approach to manipulating emotion, given a source and target data sample, is the task of converting emotion in the source sample to match the emotional style of the  target sample while retaining rest of the attributes of the source. 
While the task has shown promising results in image domain \cite{jing2019neural}, the applications in audio domain is more challenging \cite{qian2019autovc, agarwal2022leveraging}. 
In this paper, we explore the task of emotion style transfer in speech data. 

Voice conversion of speech primarily explored 
converting the speaker identity of a voice \cite{sisman2020overview}. However, speech also contains information about the underlying emotional trait of the speaker in varying levels \cite{zhou2022emotion}. The initial frameworks using Gaussian mixture model (GMM) \cite{aihara2012gmm}, hidden Markov model \cite{inanoglu2009data} and deep learning    \cite{ming2016deep} based conversion approaches have recently been advanced with 
generative adversarial networks (GAN) \cite{rizos2020stargan}  and sequence-to-sequence   auto encoding models \cite{shankar2020multi}. 

In many of the prior emotion conversion approaches, the emotion targets are treated as discrete labels. However, emotion is a fine-grained attribute which has varying levels of granularities \cite{zhou2022emotion}.
Forcing the emotion attribute to a small number of discrete target labels  may not allow the models to capture the wide range of diverse and heterogeneous sentiments elicited in  human speech. 
Hence, we argue that the most natural form of emotion conversion is to transfer the emotion expressed in a target audio to the source audio, a.k.a A2A emotion style transfer. This motivation is also echoed in a recent work on A2A style transfer~\cite{chen23_interspeech}.
 In spite of  these efforts,  audio-to-audio (A2A) style transfer in  zero shot setting (unseen speakers and emotion classes) is challenging.

On a separate front,  representation learning of speech has   shown remarkable progress in the recent years. The wav2vec \cite{baevski2020wav2vec} models have been improved with masked language modeling (MLM) objectives (for example, HuBERT \cite{hsu2021hubert}) in self-supervised learning (SSL) settings. The derivation of  speaker representations have mostly been pursued with  a supervised model \cite{desplanques2020ecapa}. 
  Further, reconstructing speech from factored representations of speaker, content and pitch contours \cite{polyak2021speech} has shown that models like Tacotron \cite{shen2018natural}, AudioLM \cite{borsos2023audiolm} and HiFi-GAN \cite{kong2020hifi} allow good quality speech generation.

In this paper, we propose a framework called, \methodexp - \method{}, which leverages the advances made in representation learning and speech reconstruction. The proposed framework decomposes the given audio into semantic tokens (using HuBERT model \cite{hsu2021hubert}), speaker representations (x-vectors \cite{desplanques2020ecapa}), and emotion embeddings (derived from a pre-trained emotion classifier). Inspired by speaker disentangling proposed for speech synthesis \cite{li2022cross}, we also perform a single step of speaker and emotion disentanglement in the   embeddings.  Since pitch (F0) is also a component that embeds content, speaker and emotion, we investigate a cross-attention based  model for predicting the F0 contour of a given utterance. 
Using the three representations (speech, speaker and emotion) along with the predicted F0 contours, the proposed \method{} framework utilizes the HiFi-GAN \cite{kong2020hifi} decoder model for reconstructing the speech. 
During emotion conversion, the proposed \method{} approach does not use text or parallel training and simply imports the emotion embedding from the target audio for style transfer.

The experiments are performed on emotion speech dataset (ESD) \cite{zhou2021seen}. 
We also explore a zero shot setting, where an unseen emotion from a different dataset is used as the reference audio. Further, a setting where the source speech is derived from an unseen speaker is also investigated.
We perform several objective and subjective quality evaluations  and compare with benchmark methods to highlight the style transfer capability of the proposed framework. The key contributions from this work can be summarized as follows,
\begin{itemize}[leftmargin=*]
    \item Proposing a novel framework for predicting the pitch contour of a given audio file  using the semantic tokens from HuBERT, speaker embeddings and emotion embeddings.
    \item Enabling speaker-emotion disentanglement  using adversarial training. 
    \item Illustrating zero shot emotion transfer capabilities from unseen emotion categories, novel speakers and content. 
\end{itemize}
\section{Related Prior Works}
\noindent \textbf{A2A EST Using World Vocoder}: One of the earliest attempts for EST involved using the world vocoder, as proposed by Gao et al. \cite{gao2018nonparallel}. This work used the statistics of F0 and spectral components from the target speaker before reconstruction using the decoder. Our work uses recent advances in  speaker, emotion and content embeddings for emotion style transfer. Further, we aim to transfer the emotion style from the reference speech to the source speech signal rather than just modifying the emotion category.

\noindent \textbf{Expressive text-to-speech synthesis}: The work by Li et al. \cite{li2022cross} explored using speaker disentanglement for generating emotional speech from text. Similarly, Emovox proposed by Zhou et al. \cite{zhou2022emotion} used phonetic transcription for emotional voice conversion. However, our work explores EST without using any linguistic or phonetic transcriptions of the source or target speech.  

\noindent \textbf{Non-parallel and unseen emotion conversion}: Recent work by Chen et al. \cite{chen23_interspeech} explored using attention models for performing EST. However, this work forced the source and target speech to be from the same speaker, limiting the utility of the EST applications. 
\section{Method}
\begin{figure}[t!]
    \centering
\includegraphics[width=0.44\textwidth,trim={0cm 14.5cm 9cm 6cm},clip]{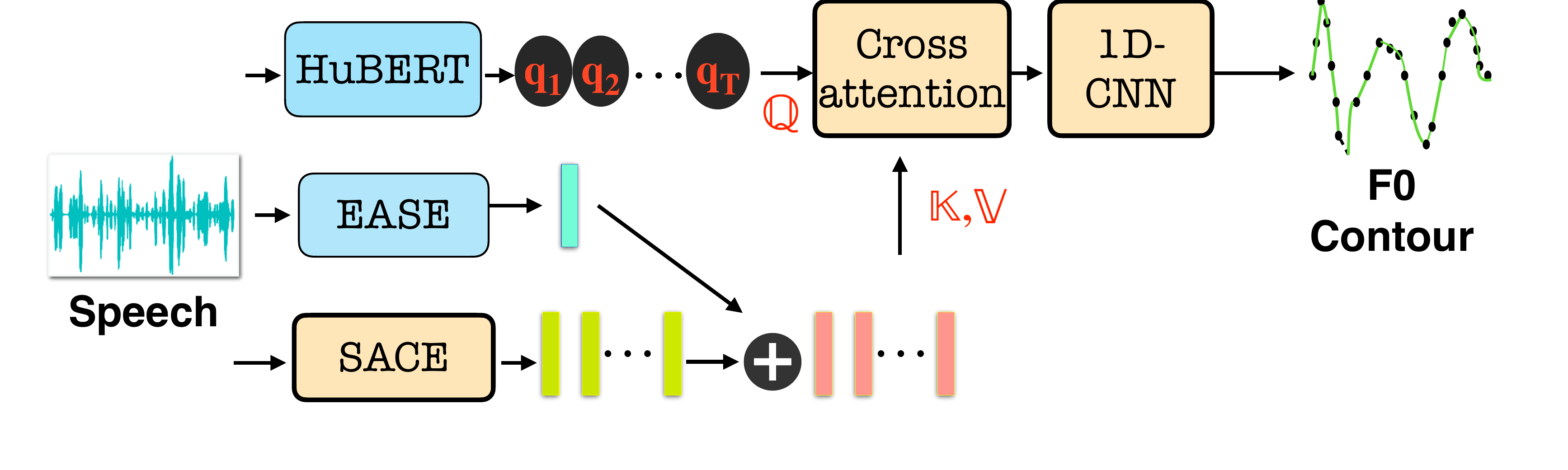}
     \vspace{-0.3cm}
    \caption{Training of the F0 contour predictor.  EASE  - Emotion Agnostic Speaker Encoder. SACE - Speaker Adversarial Classifier of Emotions.  The blue blocks are kept frozen during training.  }
    \label{fig:pitch_pred}
    \vspace{-0.4cm}
\end{figure}
\subsection{Content encoder}
\label{sec:con}
The content encoder used in the proposed framework is  the HuBERT SSL model \cite{hsu2021hubert}. 
The HuBERT model gives continuous valued vector representations for each speech segment, which is subsequently converted into discrete tokens with a k-means clustering. 
\vspace{-0.1cm}
\subsection{Emotion Agnostic Speaker Encoder (EASE)}
\label{sec:spkr}
The speaker embeddings for each audio file are extracted using a  enhanced channel attention-time delay neural network (ECAPA-TDNN)~\cite{desplanques2020ecapa} model. This model is pre-trained on $2794$ hours and $7363$ speakers from the VoxCeleb dataset~\cite{nagrani2020voxceleb}, for the task of speaker classification. 
The model involves an utterance level pooling of the frame-level embeddings, called x-vectors. 
The x-vectors have   been shown to encode emotion information~\cite{pappagari2020x, shaheen23_interspeech}. In order to suppress the emotion information,  inspired by the disentanglement approach proposed in Li et al. \cite{li2022cross}, 
we add two fully connected layers to the x-vector model and further train the model with an emotion adversarial loss~\cite{ganin2016domain}. We refer to these vectors as the Emotion Agnostic Speaker Encoder (EASE) vectors. The loss  function is given by
\begin{equation}\label{eq:lossspkr}
    \mathcal{L}_{tot-spkr} = \mathcal{L}_{ce}^{spkr} - \lambda_{adv}^{emo} \mathcal{L}_{ce}^{emo}
\end{equation}

\subsection{Speaker Adversarial Classifier of Emotions (SACE)}
\label{sec:emo}
A speaker adversarial classifier of emotions (SACE) classifier is designed based on the wav2vec2.0 representations \cite{baevski2020wav2vec}, similar to the one proposed by  Pepino et al.~\cite{pepino2021emotion}. The wav2vec model, pre-trained on $300$ hours ($543$ speakers) of switchboard corpus~\cite{godfrey1992switchboard}, is used for extracting features from the raw speech signal~\cite{dutta2023hcam}. The convolutional feature extractors are kept frozen while the transformer layers along with two position wise feed forward layers are trained for the task of emotion recognition on the Emotional Speech Dataset (ESD)~\cite{zhou2021seen}. The model is trained   with speaker adversarial loss (the emotion classifier equivalent of Eq.~\ref{eq:lossspkr}). 
The representations averaged over the entire utterance are used as the emotion embedding.
\vspace{-0.2cm}
\subsection{Pitch Contour Predictor}
\label{sec:pitchpred} 
The framework for the pitch (F0) predictor is shown in Figure~\ref{fig:pitch_pred}. 
The HuBERT tokens for the speech signal are converted to an sequence of vectors by means of an embedding layer. This sequence, denoted by $\mathbf{\mathcal{Q}} = \{q_1,..,q_T\}_{t=1}^T$, is used as the query sequence for cross-attention. The frame-level SACE embeddings are added with speaker embedding (EASE) to form the key-value pair for the cross attention module~\cite{vaswani2017attention}. This is followed by a 1D-CNN network to predict the F0 contour. The target pitch contour is the one derived using the YAAPT algorithm \cite{kasi2002yet}.   
We use the $\mathcal{L}_1$ loss between the predicted and target F0 contour.
\begin{figure}[t!]
    \centering
    \includegraphics[width=0.5\textwidth,trim={7cm 0cm 5.8cm 0cm},clip]{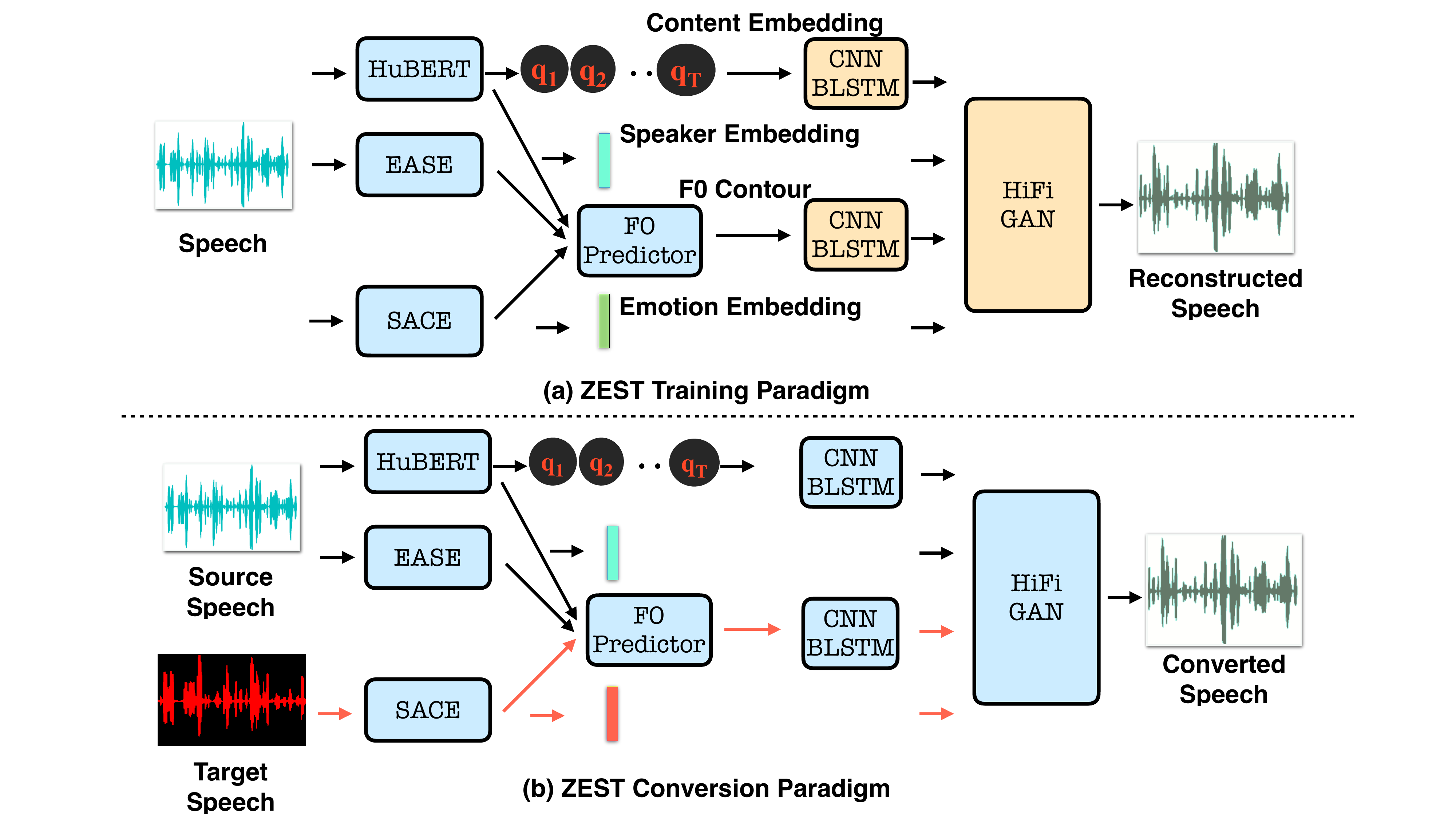}
    \caption{
    (a) During training, ZEST is learned to reconstruct the speech signal. (b) During emotion conversion, the components that are derived from target speech are coded in orange color. The yellow blocks are the  learnable parts of the model using an auto-encoding objective, while the blue blocks indicate frozen components. }
    \label{fig:model}
    \vspace{-0.5cm}
\end{figure}
\vspace{-0.1cm}
\subsection{Speech reconstruction}
\label{sec:synth}
The speech reconstruction framework is shown in Fig.~\ref{fig:model}(a). 
For reconstructing the speech signal, the HuBERT tokens, SACE embedding, EASE vector and the predicted F0 contour are used.  The HuBERT tokens are converted to a sequence of real-valued vectors with a learnable embedding layer. In order to add  contextual information during the speech reconstruction  phase, the tokens and $F0_{pred}$ are passed through two separate networks consisting of CNN layers and bidirectional long-short term memory (BLSTM) layers. Finally, all the components are passed through a HiFi-GAN vocoder~\cite{kong2020hifi} to reconstruct the speech signal. 
More details of the HiFi GAN model are provided in Polyak et al.~\cite{polyak2021speech}. 

\subsection{Emotion conversion}
\label{sec:emoconv}
The ZEST framework for emotion conversion is shown in Figure~\ref{fig:model}(b).
The HuBERT tokens and the speaker vector are extracted from the source speech  while the emotion embeddings are derived from the target speech.  
The emotion embedding sequence, being extracted from the reference speech, may differ in length from the  source speech. 
However, as the query sequence in the cross-attention (Figure~\ref{fig:pitch_pred}) is driven by the HuBERT tokens of the source signal, the F0 contour generated during conversion will match the length of the source signal.   The HuBERT tokens, speaker vector, predicted F0 contour and the emotion embedding are then used to generate the converted speech through the pre-trained HiFi-GAN model. The conversion phase does not involve any model training steps.


\section{Experiments and Results}
\subsection{Datasets and Pretraining}
\label{dataset}
The HuBERT model is the pre-trained $12$-layer base model described in Hsu et al. \cite{hsu2021hubert}. This model is pre-trained on $960$ hours of Librispeech dataset. 
The speaker encoder is initialized with the ECAPA-TDNN model \cite{desplanques2020ecapa} pre-trained on the VoxCeleb dataset. Further, the emotion agnostic training of the EASE model is performed on the Emotional Speech Database (ESD)~\cite{zhou2021seen}. We also train the F0 predictor as well as the HiFiGAN based reconstruction module on this dataset. We use the gender balanced English partition of this dataset in our experiments with $10$ speakers. The training, validation and test splits are used as suggested  in the dataset where, $300$ utterances from each of the $10$ speakers, for the five labelled emotions - neutral, angry, happy, sad and surprise, are used for training. 
Thus, the training data consists of $15000$ utterances, while $2500$ ($50$ utterances per speaker per emotion) unseen utterances are used for validation and testing. 

We further evaluate the \method{} method on unseen emotions and speakers. For unseen emotion targets, we use $2542$ utterances from the CREMA-D dataset~\cite{cao2014crema} belonging to two new emotion classes (fear and disgust). For unseen source speakers, we investigate the use of the TIMIT database~\cite{garofolo1993darpa}. We choose $100$ utterances distributed across $10$ speakers selected at random from the TIMIT dataset for these experiments.
Note that, the TIMIT and the CREMA-D datasets are only used in the conversion setting (Figure~\ref{fig:model}(b)). 
These evaluations reflect the zero-shot capability of the model. 
\vspace{-0.1cm}
\subsection{Implementation}
The HuBERT representations for the encoder are taken from  $9$-th layer of the pre-trained model and clustered with a k-means algorithm ($K=100$)~\cite{polyak2021speech}. The EASE model is trained for a total of $10$ epochs with $\lambda_{adv}^{emo}$ set to $10$ and batch size set to $32$. The SACE setup is trained with a speaker adversarial loss ($\lambda = 1000$) along with the pitch predictor (Section~\ref{sec:pitchpred}) with a batch size of $24$ and a learning rate of $1e-4$, while the speech reconstruction model is trained with a batch size of $32$ and a learning rate of $2e-4$. The pitch predictor is trained for $50$ epochs and the HiFi-GAN model is trained for $100K$ steps. The values of $\lambda_{adv}^{emo}$ and $\lambda$ are set based on the validation set performance of EASE and the pitch predictor respectively. Our code, converted audio samples and further ablations are available at this link\footnote{\url{https://github.com/iiscleap/ZEST}}.

We use the VAWGAN model proposed by Zhou et al.~\cite{zhou2021vaw} as the baseline setup for bench-marking the proposed \method{}  framework. 
As described in the paper, we supply the target emotion classes instead of the reference audio file for conversion. 
We also explore the reconstruction model proposed in Polyak et al. \cite{polyak2021speech} for benchmarking. Here, the F0 contour is derived from the target audio, while the speech and speaker contents are derived from the source audio. 

    

\subsection{Evaluation settings}\label{sec:objective-evaluation}
\begin{itemize}[leftmargin=*]
    \item \textbf{Same-Speaker-Same-Text (SSST)}: The source and reference speech are from the same speaker and with the same textual content. Further, the source speech is always in neutral emotion, while the reference speech can be from any other emotional category.
    This test set has a total of $1200$ samples ($30$ neutral samples from the $10$ speakers to be converted to $4$ other emotion categories)
    
    \item \textbf{Same-Speaker-Different-Text (SSDT)}: The source and reference speech are from the same speaker but with different textual content. This test set is created by randomly choosing $10$ neutral utterances (one per speaker) as the source speech signals, and     the remaining $29$ test set utterances for conversion into each of the four target emotions. This leads to a total of $1160$ test utterances.
    \item \textbf{Different-Speaker-Same-Text (DSST)}: The source and reference speech are derived from different speakers   using the same textual content. Each source speech signal (in neutral emotion) has $36$ reference signals (the other $9$ speakers having the same utterance in each of the $4$ emotions). As each speaker has $30$ test utterances, this test set has a total of $10800$ utterances. 
    \item \textbf{Different-Speaker-Different-Text (DSDT)}: This is the most generic setting, where the source and reference speech do not share speaker or textual content.  In this case, $10$ neutral utterances are chosen as the source speech signal set (one per speaker). For each of these utterances, the remaining $29$ utterances are selected as the reference.
    This results in a test set of $10400$ samples. 
    \item \textbf{Unseen Target Emotions (UTE)}: The $2542$ utterances from CREMA-D are considered as target speech signals for each of the $10$ randomly chosen neutral source utterances from ESD dataset, resulting in a total of $25420$ evaluation utterances. 
    \item \textbf{Unseen Source Speakers (USS)}: In this setting, the $100$ utterances from TIMIT are chosen to be the source speech signals and $8$ randomly chosen emotional utterances from ESD dataset ($2$ per emotion class barring neutral) are used as the reference speech signals. Each of the $8$ utterances from ESD is from a different speaker. This results in a total of $800$ evaluation utterances. 
\end{itemize}

\begin{table}[t!]
\caption{Objective evaluation results (\%). Here, Emo. Acc.- Emotional Accuracy, CER - Character Error Rate, Spk. Acc. - Speaker accuracy. The  different test settings are described  in Sec.~\ref{sec:objective-evaluation}.}
\label{tab:obj results}
\centering
\resizebox{0.49\textwidth}{!}{
\begin{tabular}{@{}l|l|c|c|c|c|c|c@{}}
\toprule
 & Method & SSST & SSDT & DSST & DSDT & UTE & USS \\ \midrule
\multirow{5}{*}{CER $\downarrow$} & VAWGAN~\cite{zhou2021vaw} & $7.9$ & $6.6$ & $7.9$ & $6.6$ & - & - \\ 
 & Polyak~\cite{polyak2021speech} & $\mathbf{7.0}$ & $5.7$ & $\mathbf{7.0}$ & $5.7$ & $6.1$ & $\mathbf{14.6}$ \\ \cmidrule(l){2-8} 
 & ZEST-no-adv. & $7.9$ & $\mathbf{5.6}$ & $7.6$ & $5.7$ & $6.2$ & $14.9$ \\
 & ZEST-no-F0-pred. & $8.6$ & $6.3$ & $8.5$ & $6.3$ & $7.1$ & $14.9$ \\ 
  & ZEST & $7.2$ & $\mathbf{5.6}$ & $7.2$ & $\mathbf{5.4}$ & $\mathbf{5.9}$ & $14.8$ \\ 
 \midrule
\multirow{5}{*}{\begin{tabular}[c]{@{}c@{}}Emo.\\ Acc.\end{tabular} $\uparrow$}& VAWGAN~\cite{zhou2021vaw} & $41.4$ & $44.5$ & $41.4$ & $44.5$ & - & - \\ 
 & Polyak~\cite{polyak2021speech} & $36.4$ & $33.3$ & $28.3$ & $25.9$ & - & $26.6$ \\ 
 \cmidrule(l){2-8} 
 & ZEST-no-adv. & $56.1$ & $42.4$ & $46.8$ & $38.6$ & - & $36.4$ \\
 & ZEST-no-F0-pred. & $55.8$ & $39.9$ & $45.3$ & $36.6$ & - & $28.4$ \\ 
  & ZEST & $\mathbf{69.0}$ & $\mathbf{63.8}$ & $\mathbf{60.0}$ & $\mathbf{55.6}$ & - & $\mathbf{72.5}$ \\ 
 \midrule
\multirow{5}{*}{\begin{tabular}[c]{@{}c@{}}Spk.\\ Acc.\end{tabular} $\uparrow$}& VAWGAN~\cite{zhou2021vaw} & $62.4$ & $76.0$ & $62.4$ & $76.0$ & - & - \\ 
 & Polyak~\cite{polyak2021speech} & $\mathbf{99.6}$ & $\mathbf{100}$ & $\mathbf{99.7}$ & $\mathbf{100}$ & $\mathbf{99.1}$ & - \\ 
 \cmidrule(l){2-8} 
 & ZEST-no-adv. & $98.3$ & $\mathbf{100}$ & $97.5$ & $99.8$ & $98.7$ & - \\
 & ZEST-no-F0-pred. & $96.8$ & $99.4$ & $93.4$ & $97.9$ & $97.6$ & - \\ 
 & ZEST & $99.4$ & $99.8$ & $97.8$ & $99.7$ & $98.6$ & - \\ 
 \bottomrule
\end{tabular}}
\vspace{-0.2in}
\end{table}
\vspace{-0.2cm}
\subsubsection{Objective evaluation} 
We use three objective metrics  on the converted speech signals.
\begin{itemize}[leftmargin=*]
    \item \textbf{Emotion conversion accuracy}: We evaluate the target emotion class accuracy on the converted audio using the trained emotion classifier (Sec.~\ref{sec:emo})
    \item \textbf{Textual content preservation}: The  converted speech is  passed through an automatic speech recognition (ASR) system. This ASR system used the HuBERT encoder as a pre-training module \cite{hsu2021hubert} and was trained on the supervised $960$h audio in the Librispeech \cite{panayotov2015librispeech} dataset using the CTC loss. The character error rate (CER) is computed for the converted audio with respect to the ground truth transcription of the source speech signal.  
    \item \textbf{Speaker preservation}:  We train a speaker classification model using the source speech and the ECAPA-TDNN embeddings to classify among the $10$ training speakers. A two-layer feed-forward model is employed as backend  for this task. The speaker recognition accuracy of the converted audio, with the source speaker as target, is used as measure of the speaker preservation property.
\end{itemize}
The results for these objective tests are shown in Table \ref{tab:obj results}. 
The following are the insights drawn from the objective evaluations.
\begin{itemize}[leftmargin=*]
    \item The emotion transfer accuracy for ZEST is seen to be signficantly improved over the baseline systems compared here. Even in challenging conditions, like different speakers and different text content present in the target audio, and for unseen source speakers, the ZEST is seen to perform emotion transfer effectively. 
    \item The CER results show that both the baseline system (Polyak et al. \cite{polyak2021speech} and the \method{} provide similar ASR performance for all the conditions. Even on the unseen target emotions (UTE), the \method{} is seen to preserve the CER. On the USS condition, both models show a degradation in the CER, potentially due to the mismatch in the training/test domain of  HiFi-GAN model.  
    \item The speaker accuracy of the VAWGAN is found to be inferior (SSST setting). The baseline model of Polyak et al. \cite{polyak2021speech} allows near perfect speaker classification on all settings, while the proposed ZEST also matches this performance on all settings except the DSST. However, the model based on Polyak et al.~\cite{polyak2021speech} is not seen to be effective in transfer of emotions.
    \item We show two ablations of \method~in Table~\ref{tab:obj results}, where either the emotion adversarial loss in the EASE module or the pitch predictor module is absent. The adversarial training allows disentangled representations of speaker and emotion,  which is shown to improve the objective quality results. Further, 
    the utility of EASE is seen from the improvement in the emotion accuracy across all the $6$ test settings.
    The pitch prediction module also aids the system to achieve a better ASR performance.
\end{itemize}

\begin{figure}[t!]
    \centering
    \includegraphics[width=0.5\textwidth,trim={9cm 9cm 6cm 5cm},clip]{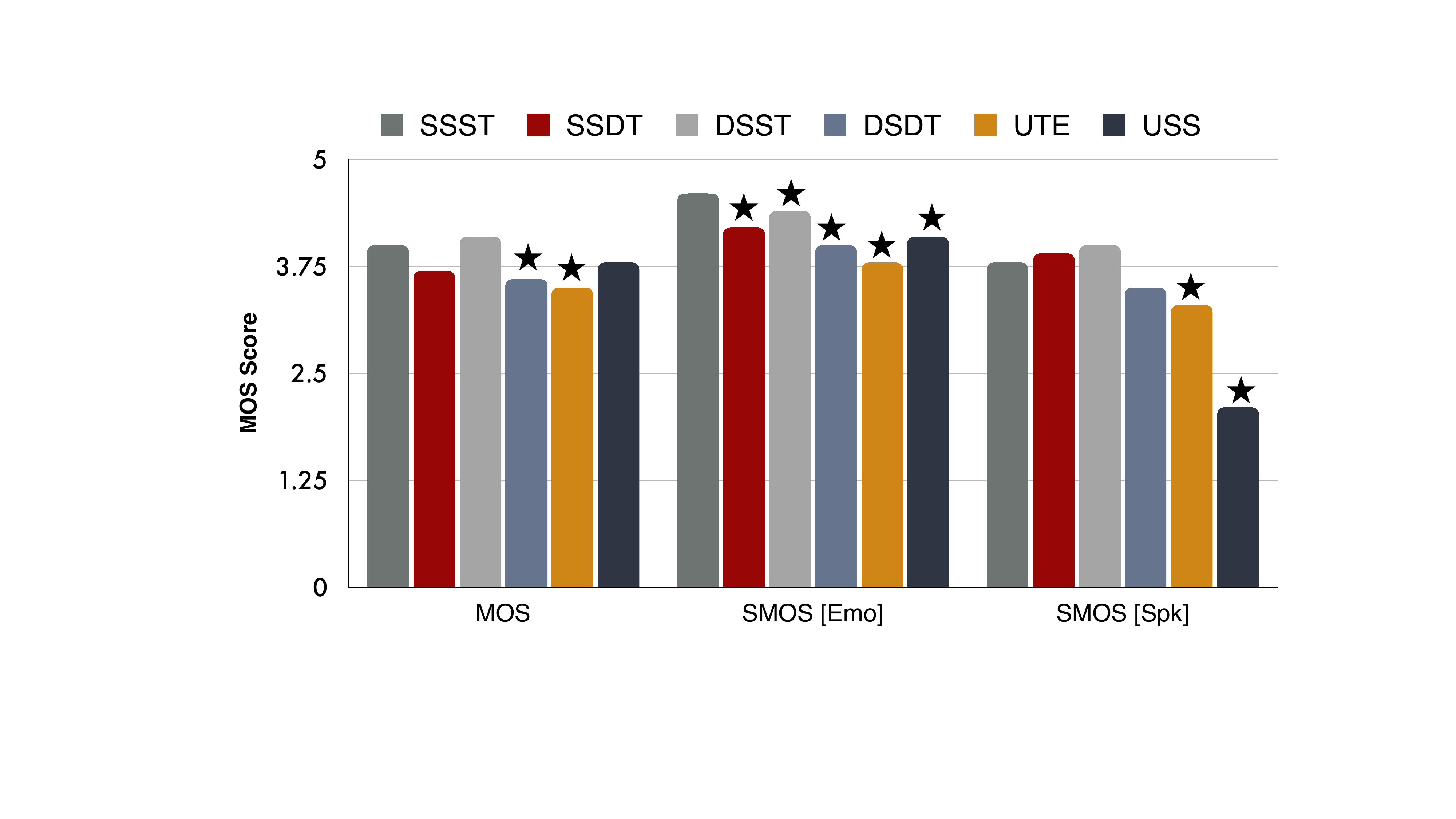}
    \caption{Subjective evaluation on the different test settings. Abbreviations used: MOS- Mean Opinion Score, SMOS - Similarity Mean Opinion Score. The definition of the different test settings is given in Sec.~\ref{sec:objective-evaluation}. {\small \FiveStar}~indicates that the difference in scores from the SSST test setting is statistically significant (p $<$ 0.05) }
     \label{fig:subjective-evaluation}
     \vspace{-0.55cm}
\end{figure}
\vspace{-0.3cm}
\subsection{Subjective tests}
We conduct listening tests in order to judge the effectiveness of our method. We used Prolific\footnote{\url{https://www.prolific.co}} tool to setup the subjective listening tests. 
We recruited $20$ participants to perform the subjective evaluation. 
We choose $52$  recordings, with $8$ recordings from each of the $4$ test settings (SSST, SSDT, DSST, DSDT) and $10$ recordings each from UTE, and USS settings. The recordings were presented in a random order. The participants were also provided with training examples to illustrate the objective of the test.
 
All the participants in the survey were asked to give their opinion score on the audio files (range of $1$-$5$) based on three criteria - i) Emotion transfer between the converted and the reference signal, ii) Quality of the converted speech and iii) Speaker similarity between the converted and the source signal. 
The subjective evaluation results (in terms of mean opinion score (MOS)) are reported in Figure~\ref{fig:subjective-evaluation}. 
As seen in these results, the emotion transfer  and reconstruction speech quality MOS values are the best for SSST condition.  This is expected as the transfer of emotion from a reference speech of the same speaker with same textual content is the easiest test setting among all. 
However, the MOS results for the other challenging conditions also compare well with the  SSST setting   (except the speaker MOS on the unseen source speaker (USS)  condition).  

We also measure a statistical significance (unpaired t-test) between the scores obtained for the SSST test setting with each of the other $5$ conditions.    
The emotion MOS for SSST is statistically significantly higher than all the other conditions.  However on the speech quality, all the conditions, except UTE and DSDT, generate statistically similar results compared to SSST setting. For the speaker MOS scores, the unseen emotion targets and unseen speaker sources generate statistically different results compared to SSST. Part of the reason for this behavior may be attributed to the limited ($10$ speakers with same speech content) training employed for the reconstruction model (HiFi-GAN). In future, we plan to train the reconstruction model on a larger resource of neutral speech along with the emotional speech to improve the generalization to novel speakers.  Further, leveraging larger emotional speech datasets for training the emotion embedding extractor and the F0 predictor may  improve the quality of the emotion transfer to unseen emotions.

\vspace{-0.15cm}
\section{Summary}
We have presented an approach for zero shot emotion style transfer (ZEST) for audio-to-audio emotion conversion. The proposed \method{}  method leverages pre-trained representations of speech content, speaker embeddings and emotion embeddings. Further, a pitch predictor model is designed in a self-supervised setting in order to learn the mapping from the representations to the pitch contour. A reconstruction module based on the HiFi-GAN allows the re-composition of the factored representations to generate the speech signal. For emotion conversion, we only derive the emotion embeddings from the target speech and perform the reconstruction. Various objective quality evaluation experiments with different levels of source/target mis-match,   transfer to unseen emotions and from unseen source speakers elicit  the zero shot emotion transfer capability of the proposed model. The subjective listening tests also validate the benefits of the proposed setting. 

\ninept
\bibliographystyle{IEEEbib}
\bibliography{refs}
\end{document}